\UseRawInputEncoding
\documentclass[%
 reprint,
 amsmath,amssymb,
 aps,
]{revtex4-2}
\usepackage{graphicx}
\usepackage{dcolumn}
\usepackage{bm}

\begin{document}

\title{Generalized Temporal Coupled Mode Theory (g-TCMT) applied to Coupled Resonator Optical Waveguides with Exchange Symmetry (CROWe)}


\author{Tianrui Li , Matthew P. Halsall and Iain F. Crowe }
\email{iain.crowe@manchester.ac.uk}
\affiliation{Department of Electrical and Electronic Engineering, Photon Science Institute, The University of Manchester, Manchester, M13 9PL, UK;}


\date{\today}

\begin{abstract}
In this paper, we extend our generalized Temporal Coupled Mode Theory (g-TCMT) model, developed earlier \citep{li_n-order_2024}, from a simple, dual-coupled micro-ring resonator (MRR) system to higher-order, $n-$serially coupled MRRs. By treating each pair of adjacent MRRs as a single `equivalent resonator', we demonstrate excellent agreement in spectral resonance position, between the g-TCMT and numerical results obtained using the transfer matrix method (TMM), for systems up to and including order $n=4$. The validity of this approach hinges on the existence, or otherwise, of \textit{exchange symmetry}, and so we refer to these structures as Coupled Resonator Optical Waveguides with exchange symmetry (CROWe's). We explore the limitations of our approach with illustrative examples of strongly coupled systems of order $n \geq 5$. Finally, we explore the properties of such higher order CROWe's for applications in non-Hermitian photonics, i.e., in which gain in some of the component MRRs and loss in the others leads to Parity-Time (PT) symmetry.
\end{abstract}


\maketitle

\section{Introduction}
The optical properties of MRRs are well established and they serve as key photonic components in a whole host of device applications, especially as part of the silicon photonics ``toolkit''  \cite{littlejohns_cornerstones_2020}. Their relatively simple design, yet diverse functionality means they turn up in devices ranging from optical modulation \cite{rabiei_polymer_2002,yuan_5_2024} to sensors \cite{dai_highly_2009,yi_highly_2010} and optical filtering \cite{prabhu_extreme_2010}. The integration of 2D materials, such as graphene and graphene oxide, has been shown to further enhance their functionality \cite{hussein_raman_2017,crowe_determination_2014,tsui_graphene_2020} and, thanks to their high Q-factor and small mode volume, they have recently been used to demonstrate on-chip nonlinear and quantum photonics \cite{leuthold_nonlinear_2010,wang_integrated_2020}. The emergence of non-Hermitian photonics in recent years, and by extension the observation of parity-time (PT) symmetry \cite{ruter_observation_2010} in coupled MRR configurations \cite{peng_paritytime-symmetric_2014,zhao_paritytime_2018} is further testament to their versatility and novel modes of operation. This new paradigm of non-Hermitian photonics has the potential to further enhance device performance, e.g., for more sensitive detection \cite{chen_exceptional_2017,ren_pt-symmetric_2018}, improved filter response \cite{zhang_bandwidth_2022}, and for on-chip laser mode-locking applications \cite{liu_integrated_2017}. Systems composed of multiple MRRs offer further  design flexibility for applications such as tunable filtering \cite{ong_ultra-high-contrast_2013,de_aguiar_automatic_2019}, optical routing \cite{ji_microring-resonator-based_2011}, and ultra-sensitive detectors \cite{youplao_microring_2018}. Recent studies have demonstrated the use of multi-MRR systems in realizing what are referred to as ``photonic molecules'' \cite{wang_fully_2021}-systems with discrete energy levels - that represent the building blocks of quantum technologies, and for performing on-chip mathematical operations, like solving differential equations \cite{sun_induced_2021,ye_reconfigurable_2024,tan_high-order_2013}, for applications in e.g., machine learning.

When multiple MRRs are coupled in series, the entire structure can be regarded as a special type of waveguide, known as a Coupled-Resonator Optical Waveguide (CROW) \cite{yariv_coupled-resonator_1999,xu_propagation_2000}. Common mathematical approaches for analyzing CROW's include Bloch's theorem \cite{yariv_coupled-resonator_1999}, and the TMM approach \cite{poon_matrix_2004}, both of which are used to elucidate properties of the guided modes in these structures, yielding, e.g., the group delay and dispersion characteristics \cite{noauthor_polymer_nodate,poon_transmission_2006}, which are important in nonlinear photonic applications. Another approach is the standard TCMT model \cite{popovic_coupling-induced_2006}, although this, like Bloch’s theorem when relies on the tight binding approximation (TBA) is only appropriate for CROW configurations with relatively weak inter-ring coupling. Furthermore, general Bloch theory is contingent on discrete translational symmetry, which is not applicable to low-order CROW systems, or to systems in which additional gain (or loss) has been introduced.

In our previous work, we developed a generalized TCMT (g-TCMT) approach, using a single, $2N \times 2N$ matrix Hamiltonian to show how one can accurately determine the eigenfrequencies of coupled MRR pairs, with arbitrarily \textit{strong} coupling \cite{li_n-order_2024}. Although the standard TCMT approach has been used to model the coupling of independent resonant modes in a single cavity \cite{wang_generating_2021}, the g-TCMT approach we developed is based on arbitrarily strong coupling between cavities. This model considers the coupling of all resonant modes of one ring, with a constant coupling rate, to the resonant modes of the other ring in a coupled MRR pair. The Hamiltonian for the coupled MRR pair (order $n = 2$) is shown in Eq. \ref{eq:1}.

\begin{equation}
    H(\Omega_{1},\Omega_{2})= \begin{bmatrix}\Omega_{1} & M \\ M & \Omega_{2} \end{bmatrix}_{2N \times 2N}
    \label{eq:1}
\end{equation}

\begin{equation*}
    \Omega_{i}=
    \begin{bmatrix}
    \ddots & \vdots & \vdots & \vdots & \vdots \\
    \cdots & \omega_{m+1} & 0 & 0 & \cdots \\
    \cdots & 0 & \omega_{m} & 0 & \cdots \\
    \cdots & 0 & 0 & \omega_{m-1} & \cdots \\
    \vdots & \vdots & \vdots & \vdots & \ddots
    \end{bmatrix}_{N\times N}-j\gamma_{i} I_{N\times N}
\end{equation*}
\begin{equation*}
    M=-\mu J_{N\times N}
\end{equation*}

Here $\gamma_{i=1,2}$ is the loss (or gain) rate in rings 1, 2, and $\omega_{m-1}$ is the lower frequency neighbouring resonance of $\omega_{m}$, with the free spectral range (FSR) given by: $\Delta\omega=|\omega_{m-1}-\omega_{m}|$. The dimension, $N$ is representative of the number of modes under consideration for each ring, $J$ is the unit matrix and $I$ is the identity matrix, so $\Omega_{i}$ is diagonal. The coupling rate between the two rings, $\mu$ is given by Eq. \ref{eq:2}, as a function of the transmission $t_c$ and coupling $k_c$ coefficients, with $t_{ri}$ the round-trip time in each ring and for lossless coupling, $k_c^2+t_c^2=1$.

\begin{align}
    \label{eq:2}
    \mu=&\frac{2k_c}{1+t_c}\frac{1}{\sqrt{t_{r1}t_{r2}}}\\=&2\tan(\frac{\arccos (t_{c})}{2}) \frac{1}{\sqrt{t_{r1}t_{r2}}} \text{, } t_{c}\in(-1,1) \notag
\end{align}

Whilst Bloch theory is convenient for analysing long chains of coupled micro-rings, enabling the study of dispersion relations, group velocity, and lattice-effects, it is limited to translational symmetry. Standard TCMT, which has also been widely adopted for modelling coupled resonators, and especially in studying PT-symmetric systems, provides a compact Hamiltonian description that facilitates the interpretation of resonance spectra in terms of eigenvalues and eigenstates. However, it is also limited to the weak coupling regime and is only valid over a narrow spectral range (viz single resonance).

On the contrary, the transfer matrix method (TMM) is capable of dealing with strongly coupled resonators. However, as it originates from classical electrodynamics and is primarily formulated in terms of field propagation and scattering, it does not naturally yield a reduced Hamiltonian or eigenvalue problem, compared with, say TCMT, making it less suitable for analysing non-Hermitian phenomena that rely explicitly on modal eigenvalue coalescence, such as in PT symmetric systems exhibiting exceptional points.

Taken together, existing theoretical approaches do not provide a unified and practical framework for describing non-Hermitian phenomena in strongly coupled optical microcavities exhibiting comb-like spectra. This limitation has hindered systematic exploration of non-Hermitian physics in micro-ring systems, especially where strong coupling is a key feature. To address this gap, we previously proposed the generalized TCMT (g-TCMT) framework for coupled ($n = 2$) resonators, and in this contribution we show how this can be applied successfully for a specific subclass of higher order ($n > 2$) MRR systems.

Following the standard TCMT approach, but extending this to our g-TCMT model we can determine the Hamiltonian, $H$ of higher order CROW systems, e.g., for an order $n=3$ system, $H$ can be written as Eq. \ref{eq:3}, where $M_{13}$ is a zero block.

\begin{align}
    \label{eq:3}
    H=\begin{bmatrix}\Omega_{1} & M_{12} &M_{13} \\ M_{12} & \Omega_{2} & M_{23}  \\ M_{13} & M_{23} & \Omega_{3} \end{bmatrix}_{3N \times 3N}
\end{align}

\begin{figure}[h]
\centering\includegraphics[width=1\linewidth]{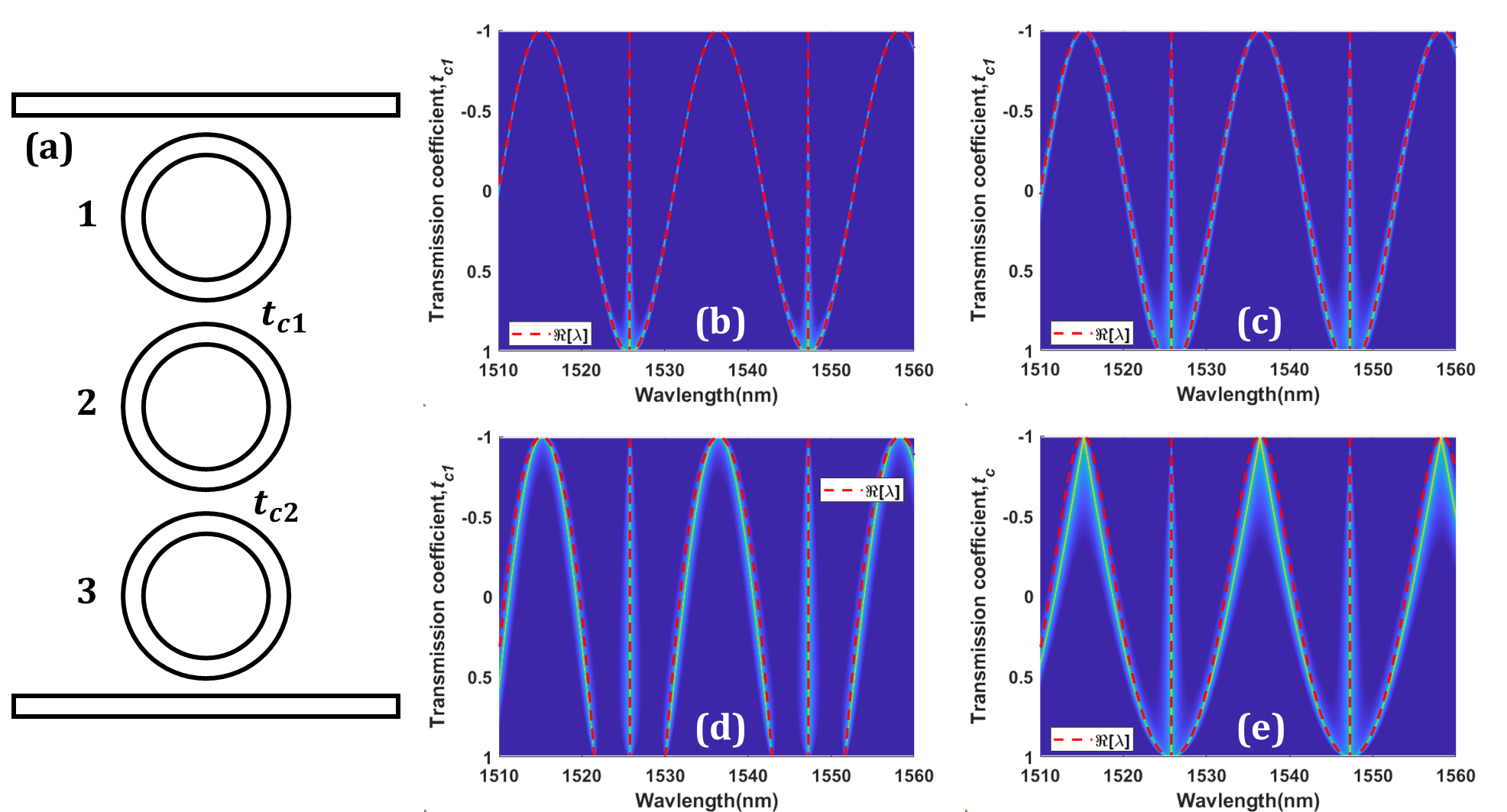}
\caption{(a) Symmetric order $n = 3$ CROW and corresponding spectral response, by TMM (colour map) and g-TCMT eigenvalues (red dashed lines) for (b) $t_{c2}=0.99$, (c) $t_{c2}=0.91$, (d) $t_{c2}=0.3$, (e) $t_{c2}=t_{c1}$.} \label{fig:-1}
\end{figure}

For the special case when the three rings are identical, as shown in Fig. \ref{fig:-1} (a), the evolution of spectral resonances derived from the eigenvalues of Eq. \ref{eq:3}, agrees rather well with that obtained from TMM, over the entire range of coupling strength, Fig. \ref{fig:-1} (b-e).

\begin{figure}[h]
\centering\includegraphics[width=1\linewidth]{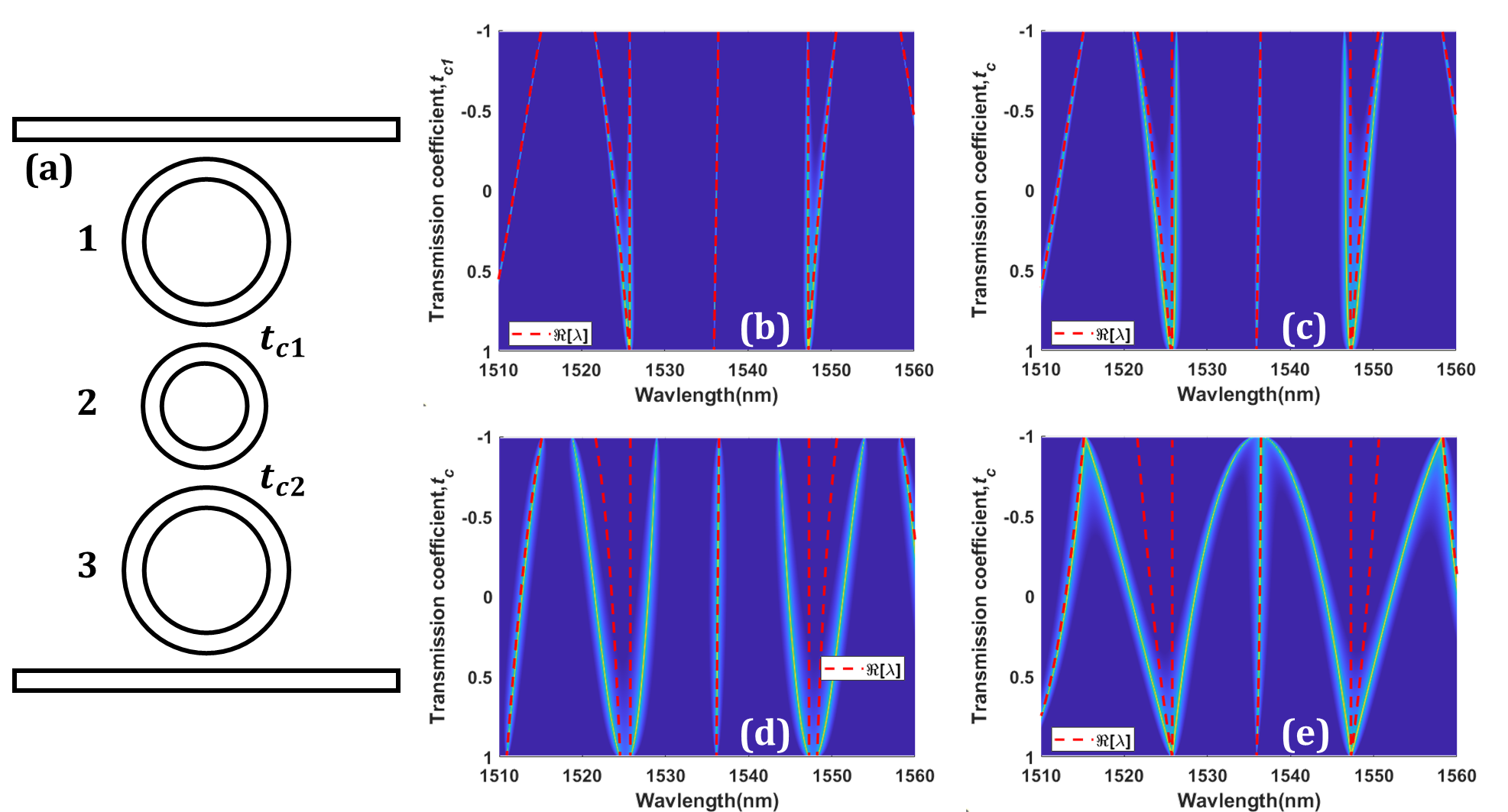}
\caption{(a) Asymmetric order $n = 3$ CROW and corresponding spectral response, by TMM (colour map) and g-TCMT eigenvalues (red dashed lines) for (b) $t_{c2}=0.99$, (c) $t_{c2}=0.91$, (d) $t_{c2}=0.3$, (e) $t_{c2}=t_{c1}$.} \label{fig:0}
\end{figure}

However, we note that, when the coupling strength is large for both inter-ring gaps, especially when $-1<t_c<0$, the numerical results derived from the g-TCMT model begin to deviate from that derived from TMM, as illustrated in Figs. \ref{fig:-1} (d, e). This breakdown in our g-TCMT model is exacerbated further in higher order asymmetric CROW systems, i.e., those with unequal MRR geometries, as shown in Fig. \ref{fig:0} (a).

Again, Fig. \ref{fig:0} (b, c), reveals rather good agreement between the spectral resonance positions derived from our g-TCMT model and that derived from TMM, but this is not the case in Fig. \ref{fig:0} (d, e) for strong coupling, where there is poor reproducibility of some of the spectral resonances as $t_c$ evolves. We conclude from this that it is not possible to provide a complete description of the spectral response of higher order ($n \geq 3$), especially asymmetric, CROW systems with arbitrary inter-ring coupling strength, using a single matrix formalism within our g-TCMT model.

The reason for this stems from the assumption that $M_{13}$ in Eq. \ref{eq:3} is a zero block, which implies zero coupling between the $1^{st}$ and $3^{rd}$ micro-rings. Whilst this is a common assumption in standard TCMT, from the perspective of classical electrodynamics, the $2^{nd}$ (middle) micro-ring may be regarded as a waveguide itself, enabling coupling between the 1st and 3rd micro-rings, and thus the assumption that $M_{13} = 0$ is not necessarily valid.

The standard TCMT represents an approximate form of g-TCMT when the block dimension is set to one. However, should the $M_{13}$ block be non-zero, the two models become irreconcilable, yielding contradictions. Consequently, neither the TCMT nor its extended form, g-TCMT, can adequately describe this coupling process.

\section{A novel g-TCMT implementation for higher order CROW’s}

As mentioned above, the limitation of this approach is that it is not possible to derive the eigenfrequencies of a serially coupled MRR array of higher orders ($n \geq 3$) using the single matrix formulism of Eq. \ref{eq:3}. However, based on a physical assumption that two coupled resonators may be regarded as a single resonator with identical resonant frequencies, by treating (some of the) coupled MRR pairs in a higher order CROW array, as single `equivalent resonators', with round-trip time, $t_{r(i_1+i_2)}$ being the sum of that of the individual component MRRs, and with appropriate considerations of the phase, and inter-ring transmission coefficient, $t_c$, we can iteratively deduce the eigenfrequencies (spectral resonances) for the entire higher order CROW. We do this by mathematical transposition of the original MRR pair, and, as such we prefer to use the term `equivalent resonator', represented using rectangles in figures from here on in. It is important to note that these are not `equivalent MRRs' because they lack some of the key attributes of what constitutes a true single MRR device, i.e., comb spectrum of equidistant single resonances. Rather these `equivalent resonators' serve as a purely mathematical route through which we can retrieve the spectral response of higher order CROW systems, with arbitrarily strong inter-coupling, by iteratively solving Eq. \ref{eq:1}.

To illustrate the method, we consider example systems consisting of order $n = 3$ and 4 serially coupled MRRs. The transmission coefficients, $t_c$, between adjacent rings are assumed to be identical and variable. Other parameters we use to demonstrate this are listed in Table \ref{tab:1}, including the guided mode effective index, $n_{eff}$, which we derive from the assumption of a waveguide core index of 3.48 (for silicon)  and a cladding index of 1.44 (for $SiO_2$) \citep{dong_novel_2014}.

\begin{table}[h] 
\caption{Fixed model parameters used to derive the results. \label{tab:1}}
\begin{ruledtabular}
\begin{tabular}{lcdr}
Parameter &  value\\
\colrule
Radius of each ring $R$ & $6.2 \mu m$ \\
Input bus-to-lower ring and upper ring-to-output\\ bus coupling coefficients, $k_{in}$ and $k_{out}$  & 0.1\\
Effective refractive index $n_{eff}$  & 2.82
\end{tabular}
\end{ruledtabular}
\end{table}

\begin{figure}[h]
\centering\includegraphics[width=1\linewidth]{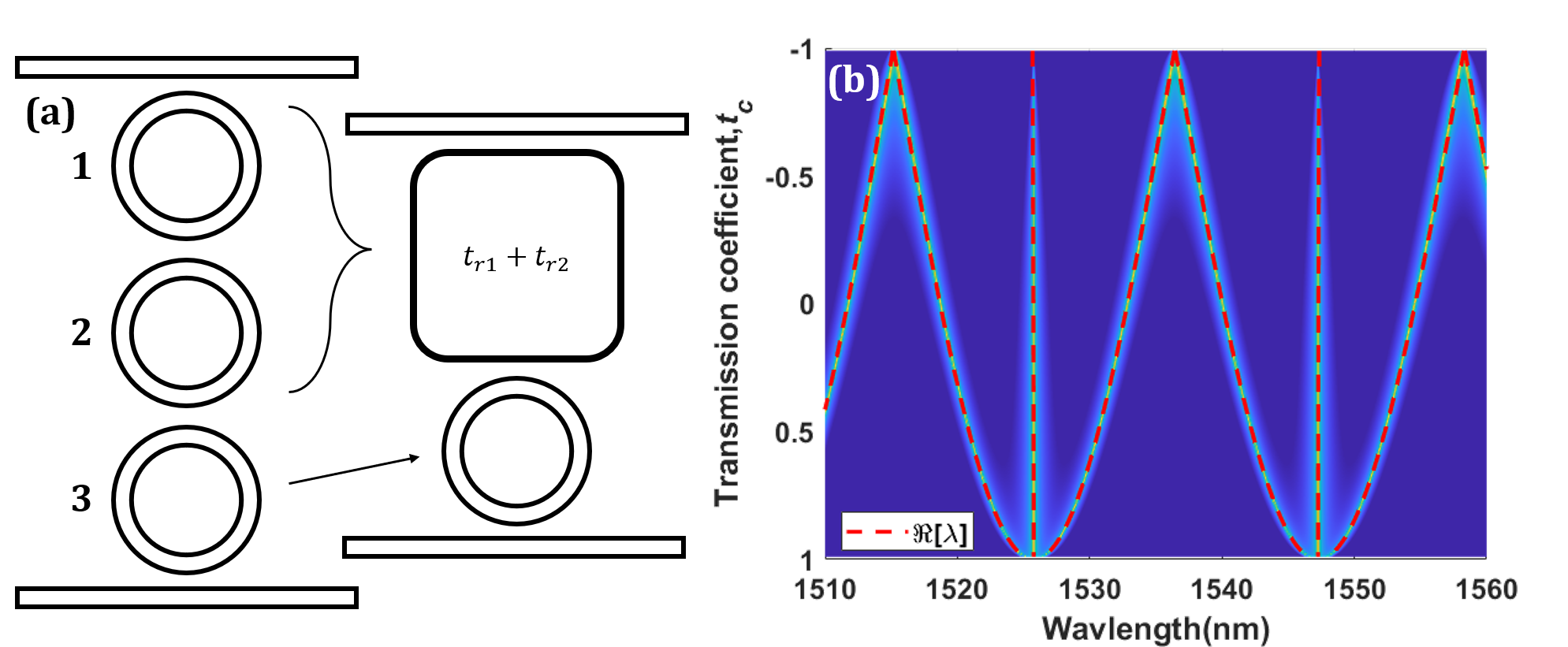}
\caption{(a) System transposition from an identical, order $n = 3$ CROW to an equivalent resonator pair (b) spectral response by TMM (colour map) and g-TCMT eigenvalues (red dashed lines).} \label{fig:1}
\end{figure}

Fig. \ref{fig:1} (a) shows the schematic model of a serially coupled, order $n = 3$ CROW consisting of 3 identical MRRs,  as in Fig. 1 (a), but modified here to show the transposition to an equivalent resonator pair. The spectral response (resonance wavelength as a function of transmission coefficient, $t_c$), again obtained using both the TMM approach and, directly, from the eigenvalues of our g-TCMT model, is shown in Fig. \ref{fig:1} (b).

For the CROW system of Fig. \ref{fig:1}, the transposition of the upper MRR pair ($R_1$ and $R_2$) to a single, larger equivalent resonator involves computing the eigenvalues, $\Omega_{12}^{'}$, of the matrix of $H(\Omega_1,\Omega_2)$, Eq.(1) as a function of the inter-ring transmission coefficient, $t_c$. This `equivalent resonator' is then coupled to the remaining MRR from the original array ($R_3$), and the eigenvalues computed again, this time for the matrix, $H(\Omega_{12}^{'},\Omega_3)$, to obtain the resonant frequencies of the original order $n = 3$ CROW. Comparison of the resonance spectra derived this way, with that derived via TMM, Fig. \ref{fig:1} (b) reveals excellent agreement between the two approaches over the entire range of $t_c$ .

\begin{figure}[h]
\centering\includegraphics[width=1\linewidth]{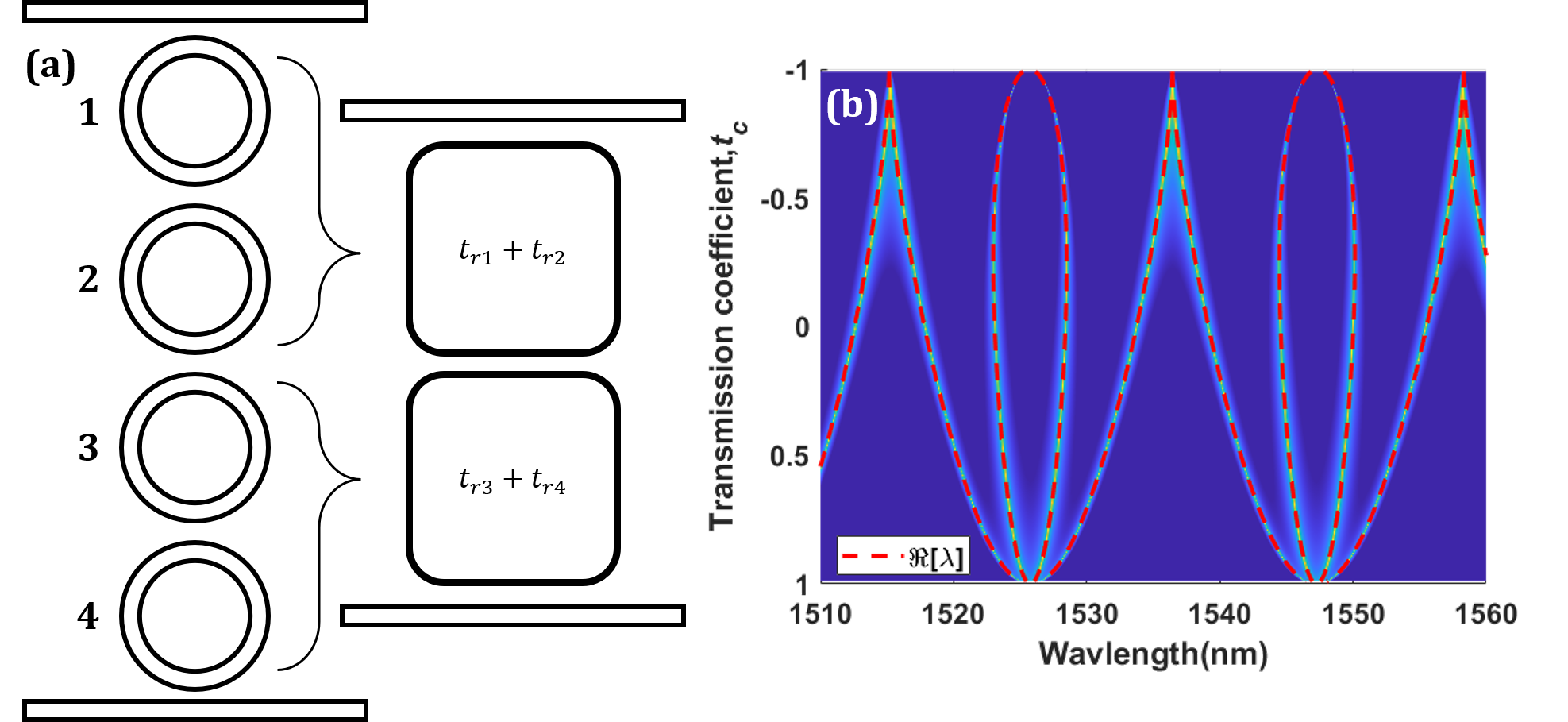}
\caption{(a) g-TCMT transposition from an identical, order $n = 4$ CROW to an identical equivalent resonator pair (b) spectral response by TMM (color map) and g-TCMT eigenvalues (red dashed lines).} \label{fig:2}
\end{figure}

Similarly, Fig. \ref{fig:2} shows the results for a symmetric order $n = 4$  CROW, i.e., consisting of 4 identical MRRs. In this case, both the upper and lower MRR pairs of the system, $R_1$, $R_2$ and $R_3$, $R_4$ are transposed to a single pair of larger `equivalent' resonators, yielding the overall resonant frequencies of the full order $n = 4$ CROW, for which the spectral response is again verified, and is in excellent agreement with the TMM result.

\section{Limitations of the approach}

\begin{figure}[h]
\centering\includegraphics[width=\columnwidth]{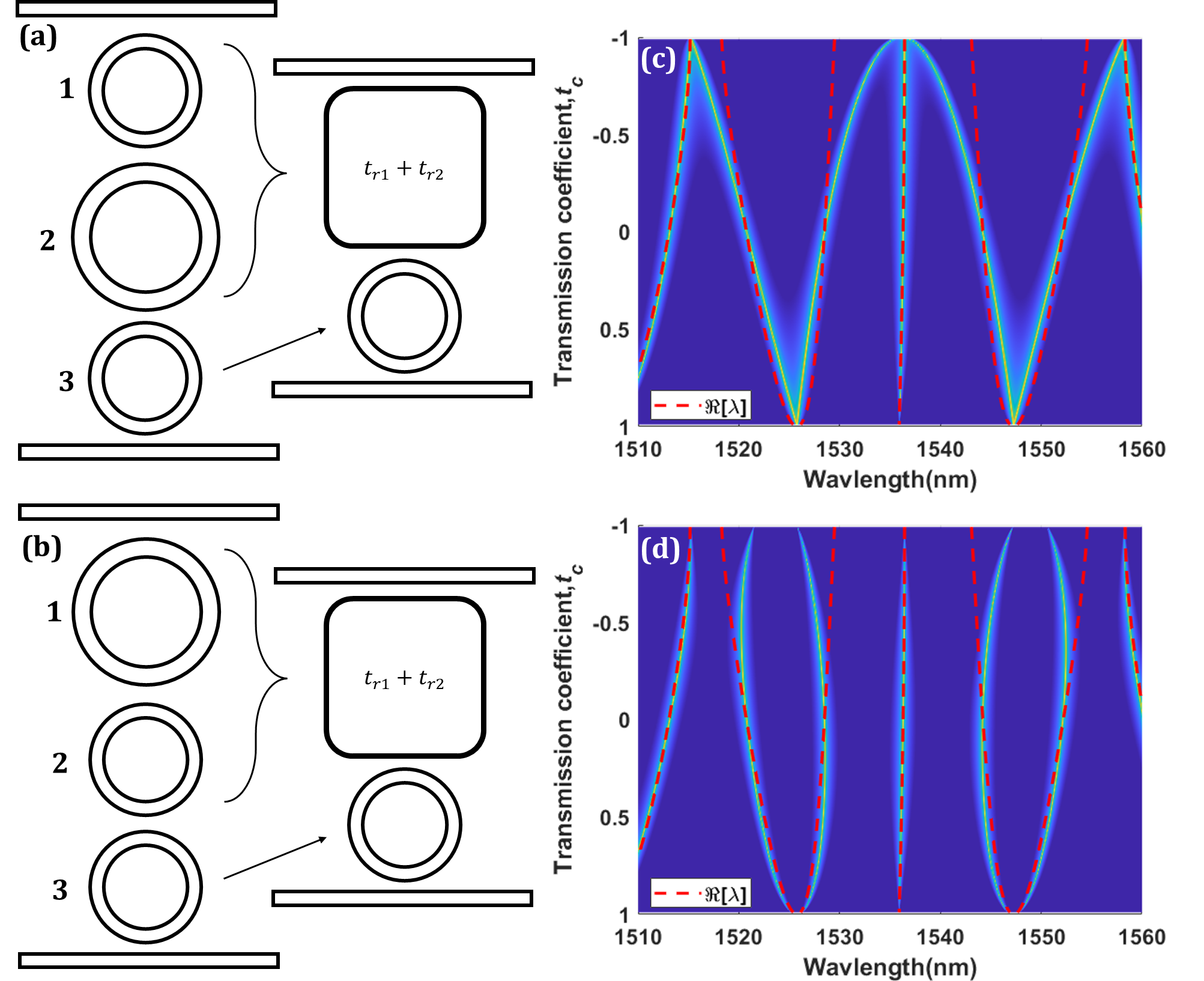}
\caption{Asymmetric order $n = 3$ CROW's with different (a) middle and (b) upper MRR. Spectral response by TMM (color map) and g-TCMT eigenvalues (red dashed lines) (c,d) corresponding to the order $n = 3$ CROW structures of (a) and (b).} \label{fig:3}
\end{figure}

In principle, this process may be applied by iterative reduction of any number of component-MRR pairs, to obtain the spectral features of such higher order systems. So far, we have shown that this approach works for initial CROW systems comprised of identical (symmetric) MRRs, but to check if this can indeed be generalized to any such higher order MRR array, we first examine the effect of MRR asymmetry and ordering in the initial CROW system. For example, Fig. \ref{fig:3} (a) and (b) show two different order $n = 3$ asymmetric CROW's, where either the middle MRR or upper MRR differ in size from the other two. 

In this case, the spectral response of the different order $n = 3$ asymmetric CROW's are expected to be different, and this is confirmed by the corresponding TMM-derived spectra shown in Fig. \ref{fig:3} (c) and (d). However, we find that in this case the iterative g-TCMT approach fails to discriminate between the different structures, nor do the eigenvalues derived from it accurately reflect the evolution of the resonances with coupling, especially in the regime of strong (over-) coupling, $0 \geq t_c \geq -1$.

\begin{figure}[h]
\centering\includegraphics[width=\columnwidth]{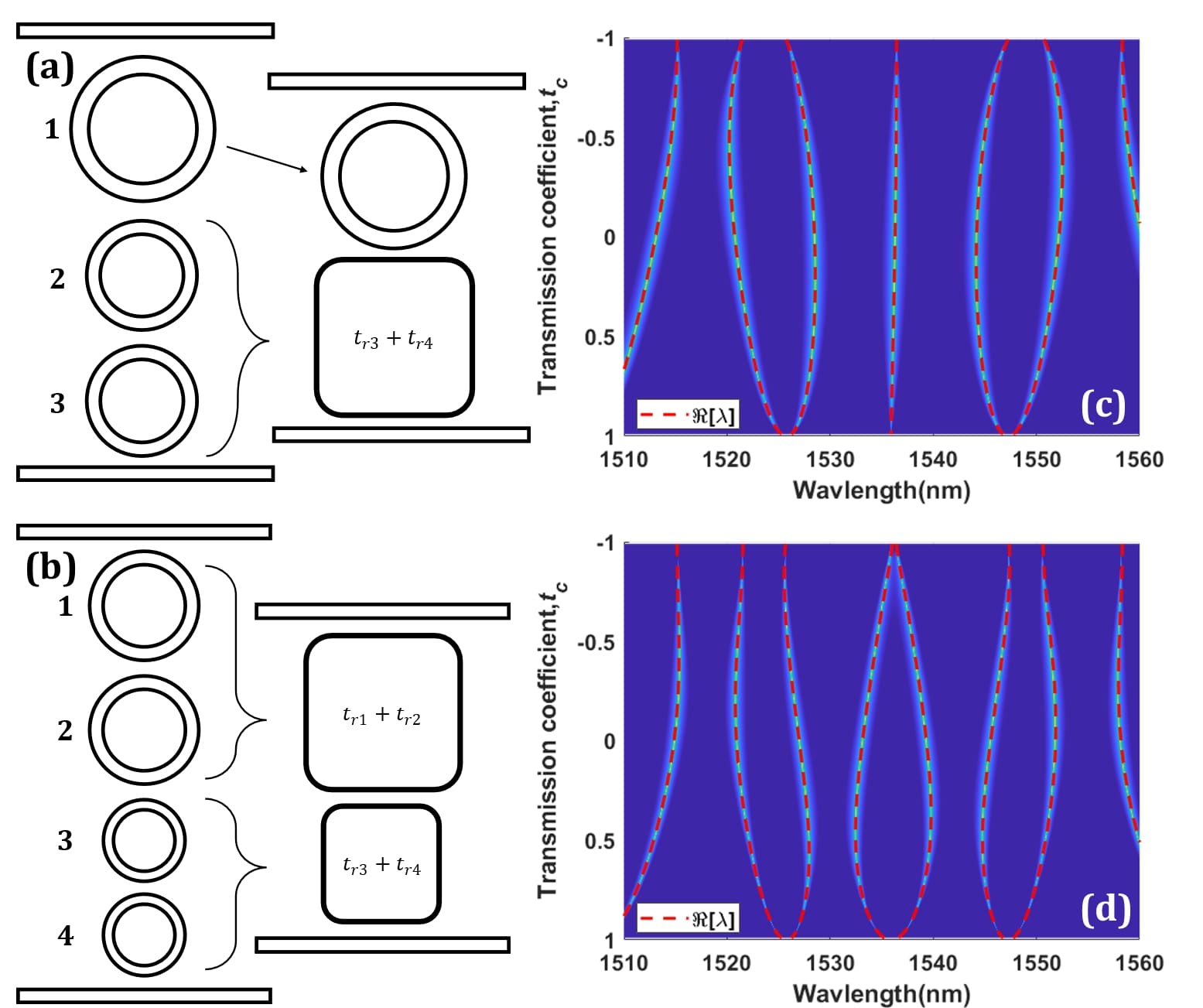}
\caption{(a) Order $n = 3$ and (b) order $n = 4$ asymmetric CROWe structures (with at least one MRR pair exhibiting exchange symmetry), transposed to an equivalent resonator pair, and their corresponding spectral response (c) and (d) by TMM (colour map) and g-TCMT eigenvalues (red dashed lines)} \label{fig:5}
\end{figure}

In fact, the g-TCMT approach is only applicable in such higher order CROW's when transposed pairs of MRRs, including the `equivalent resonators' formed at intermediate steps, satisfy what we refer to as \textit{exchange symmetry}. That is, the transposition of any resonator pair, to an equivalent resonator, must satisfy the condition that the overall system remains unchanged. A mathematical representation of exchange symmetry, based on group theory is provided in Appendix A. This concept is analogous to Bosonic \textit{exchange symmetry} in quantum field theory whereby the wave function of a system of identical bosons is symmetric under the exchange of any two particles and so remains unchanged. A subtle difference is that Bosonic \textit{exchange symmetry} applies to any particle within the system, whereas \textit{exchange symmetry} as defined in our model necessarily applies only to the transposition of adjacent resonators. As such, we refer to such structures as Coupled-Resonator Optical Waveguides with \textit{exchange symmetry} (CROWe's), but it is also important to point out that CROWe's do not require all the component MRRs of the higher order array to be identical, viz. the systems shown in  Fig. \ref{fig:5}.

In Fig. \ref{fig:5} (a), whilst $R_1$ and $R_2$ do not satisfy exchange symmetry, $R_2$ and $R_3$ do. In this case the transposition order matters, and this system may be treated as a CROWe if and only if the transposition order considers those rings with exchange symmetry ($R_2$ and $R_3$) in the intermediate step, first. This also works for order $n = 4$ CROWe systems such as that shown in Fig. \ref{fig:5} (b), where the upper MRR pair ($R_1$ and $R_2$) are different from the lower MRR pair ($R_3$ and $R_4$). In this case, $R_1$, $R_2$ and $R_3$, $R_4$ are exchange symmetric, i.e., $R_1 = R_2 \neq R_3 = R_4$ and for CROWe's transposed in this way, we find that the g-TCMT-derived spectral response, Fig. \ref{fig:5} (c) and (d) provide an excellent match to that derived by the TMM. 

\begin{figure}[h]
\centering\includegraphics[width=0.75\linewidth]{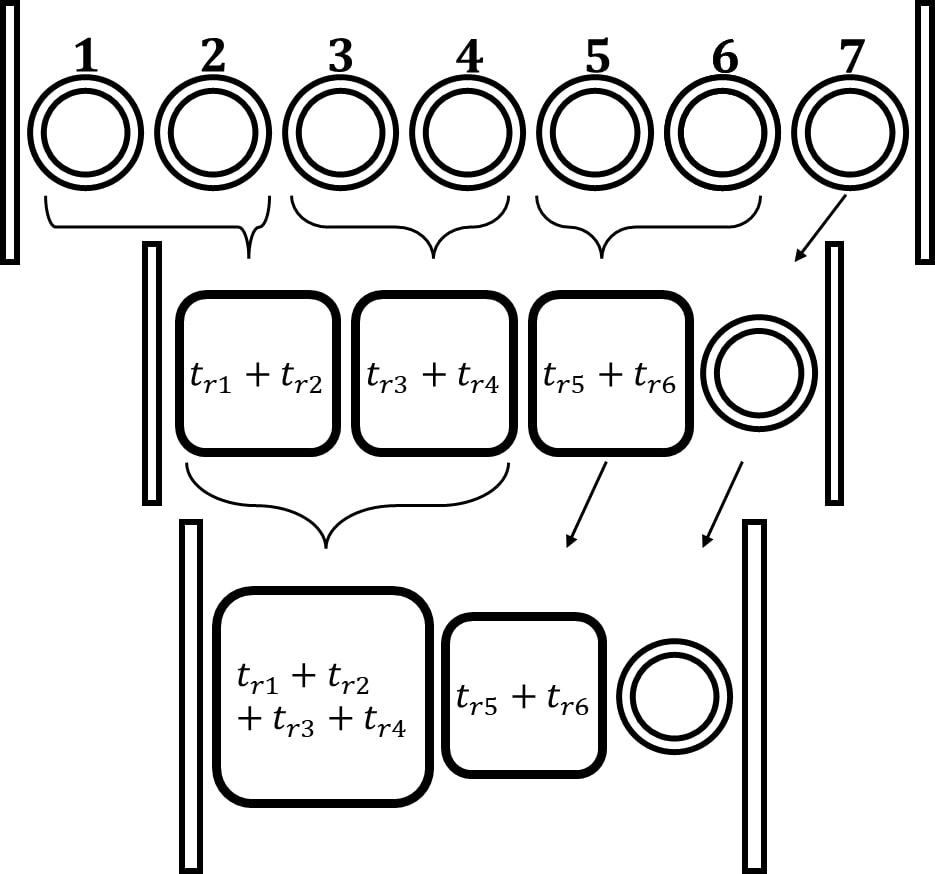}
\caption{ The 2-step transposition of an order $n = 7$ symmetric CROW system leading to an equivalent order $n=3$ resonator system, for which the spectra cannot be obtained directly from the g-TCMT approach.} \label{fig:6}
\end{figure}

Conversely, higher order CROW structures composed only of identical MRRs do not necessarily satisfy the requirements of a CROWe system, as exemplified by the order $n = 7$ symmetric CROW configuration shown in Fig. \ref{fig:6}. 

For this system, after the $1^{st}$ iteration, although the intermediate step contains a single exchange symmetric pair, from which the left-most equivalent resonator in the final step is derived, the final equivalent set of resonators is of order $n = 3$, none of which satisfy exchange symmetry. Obviously, Fig. \ref{fig:6} represents just one of many possible transposition routes to the final equivalent resonator system, and we could just as easily have taken the $6^{th}$ and $7^{th}$ rings as our starting point. In fact, there are no possible transposition routes for this CROW that yields an exchange symmetric resonator pair in the $(i - 1)^{th}$ iteration. As such this system does not constitute a CROWe and so its resonance spectrum cannot be retrieved using our g-TCMT model.

\begin{figure}[h]
\centering\includegraphics[width=\columnwidth]{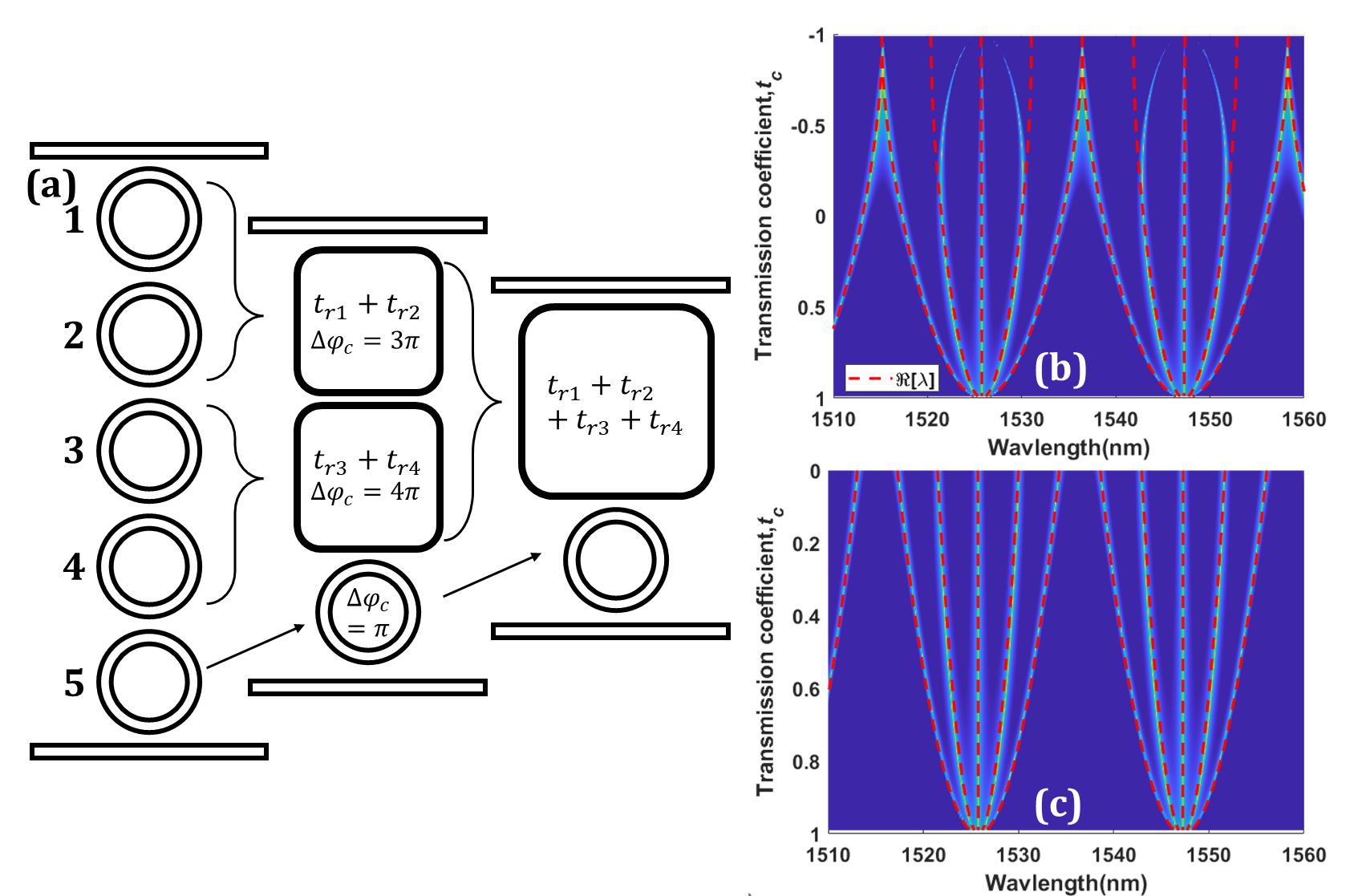}
\caption{ (a) 2-step transposition of an order $n = 5$ symmetric CROW and corresponding spectral response in the range (b) $-1 \leq  t_c  \leq  1$ and (c) $0 \leq t_c  \leq  1$ by TMM (colour map) and g-TCMT eigenvalues (red dashed lines).} \label{fig:7}
\end{figure}

For CROW structures of order $n > 4$, for this intermediate step, states arising during the successive transposition of MRRs to the final equivalent resonator pair, phase shifts that result from over-coupling ($-1 \leq t_c \leq 0$) are what prevent the equivalent resonators from satisfying the CROWe condition. Consequently, g-TCMT cannot reliably reproduce the spectral response in this strong coupling regime, for systems with order $n > 4.$ To illustrate this limitation, we consider examples of order $n = 5$ and $n = 6$ CROW systems. Fig. \ref{fig:7} (a) illustrates the 2-step transposition of an order $n = 5$, symmetric CROW.

For coupled MRRs, in the strong (over-) coupling regime ($-1 \leq t_c \leq 0$), inter-ring coupling introduces a phase shift, $\Delta \varphi_{c}=\pi$, the origin of which we have described in our earlier work \cite{li_n-order_2024}. This results in a phase imbalance for the three equivalent resonators derived in the intermediate state, i.e., $\Delta \varphi_{c}=3 \pi$ (for $R_1$ and $R_2$), $\Delta \varphi_{c}=4\pi$ (for $R_3$ and $R_4$) and $\Delta \varphi_{c}=\pi$ (for $R_5$), which in this case is what prevents them from being truly exchange symmetric, in this range of $t_c$. Notice that, in this case, this is true even though the final step apparently leads to an equivalent resonator `pair'. Consequently, the spectral response of this system in the strong coupling regime diverges from the TMM result, as shown in Fig. \ref{fig:7} (b). In contrast, in the weak (under-) coupled regime, i.e., for $0\leq t_c \leq 1$, no such relative phase differences exist between constituent resonators in the intermediate state and so they maintain exchange symmetry in this range of $t_c$ and the system meets the CROWe requirements, as exemplified by the excellent agreement between the g-TCMT and TMM derived spectral response shown in Fig. \ref{fig:7} (c).

\begin{figure}[h]
\centering\includegraphics[width=\columnwidth]{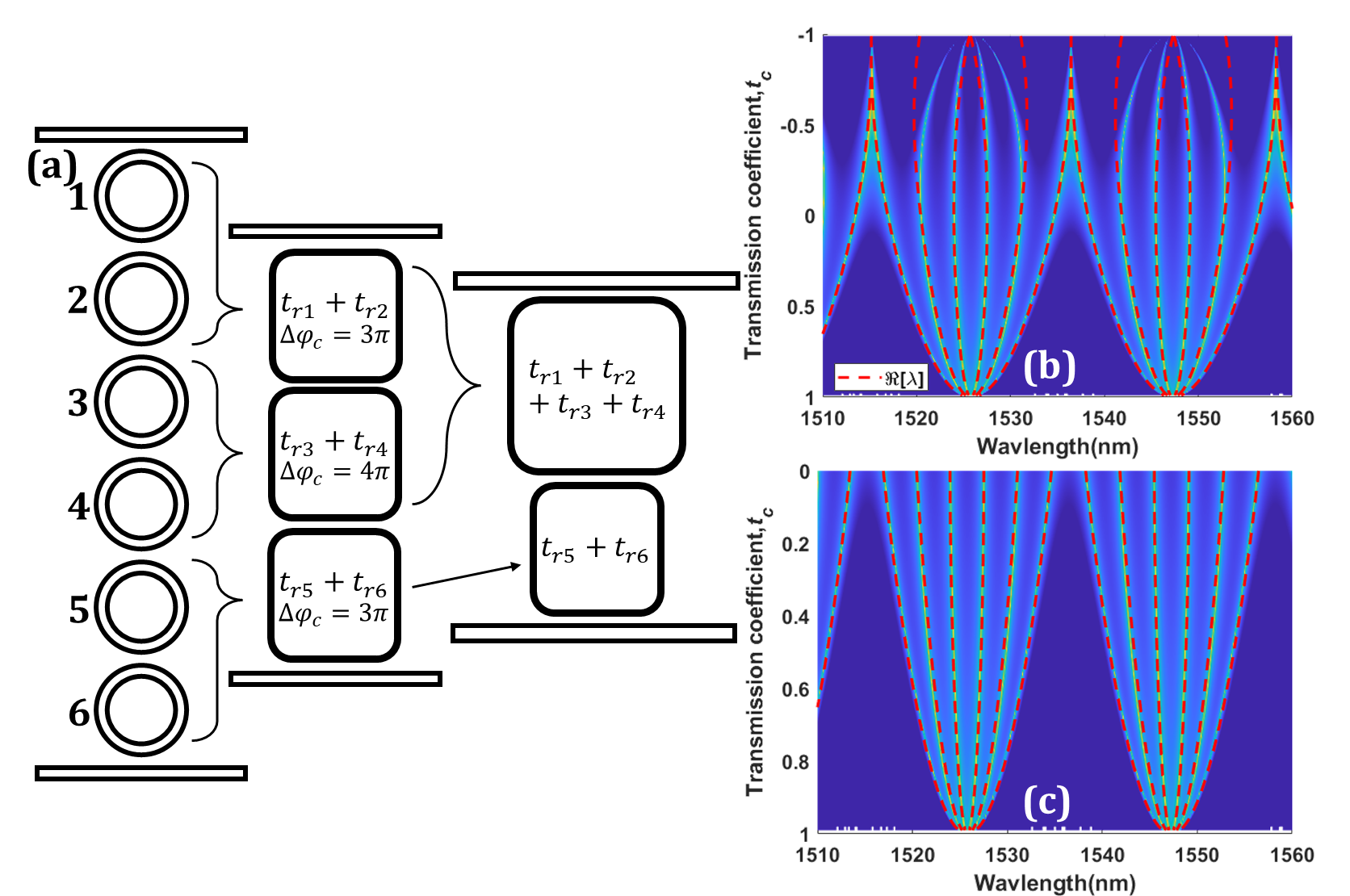}
\caption{ (a) 2-step transposition of an order $n = 6$ symmetric CROW and corresponding spectral response in the range (b) $-1 \leq  t_c  \leq  1$ and (c) $0 \leq t_c  \leq  1$ by TMM (colour map) and g-TCMT eigenvalues (red dashed lines).} \label{fig:8}
\end{figure}

A similar behavior is observed for an order $n = 6$ symmetric CROW, in which the intermediate step retains exchange symmetry only under the conditions of weak coupling, Fig. \ref{fig:8}.

In the case of the CROW system of Fig. \ref{fig:8}, the phase imbalance for the three equivalent resonators in the intermediate step, $\Delta \varphi_{c}=3 \pi$ (for $R_1$ and $R_2$), $\Delta \varphi_{c}=4 \pi$ (for $R_3$ and $R_4$) and $\Delta \varphi_{c}=3 \pi$ (for $R_5$ and $R_6$) under the conditions of strong coupling ($-1 \leq t_c \leq 0$) again breaks the exchange symmetry, leading to poor agreement between the g-TCMT and TMM derived spectral responses in this range.

\section{A non-Hermitian, order $n = 4$ CROWe exhibiting Parity-Time (PT) symmetry}

\begin{figure}[h]
\centering\includegraphics[width=\columnwidth]{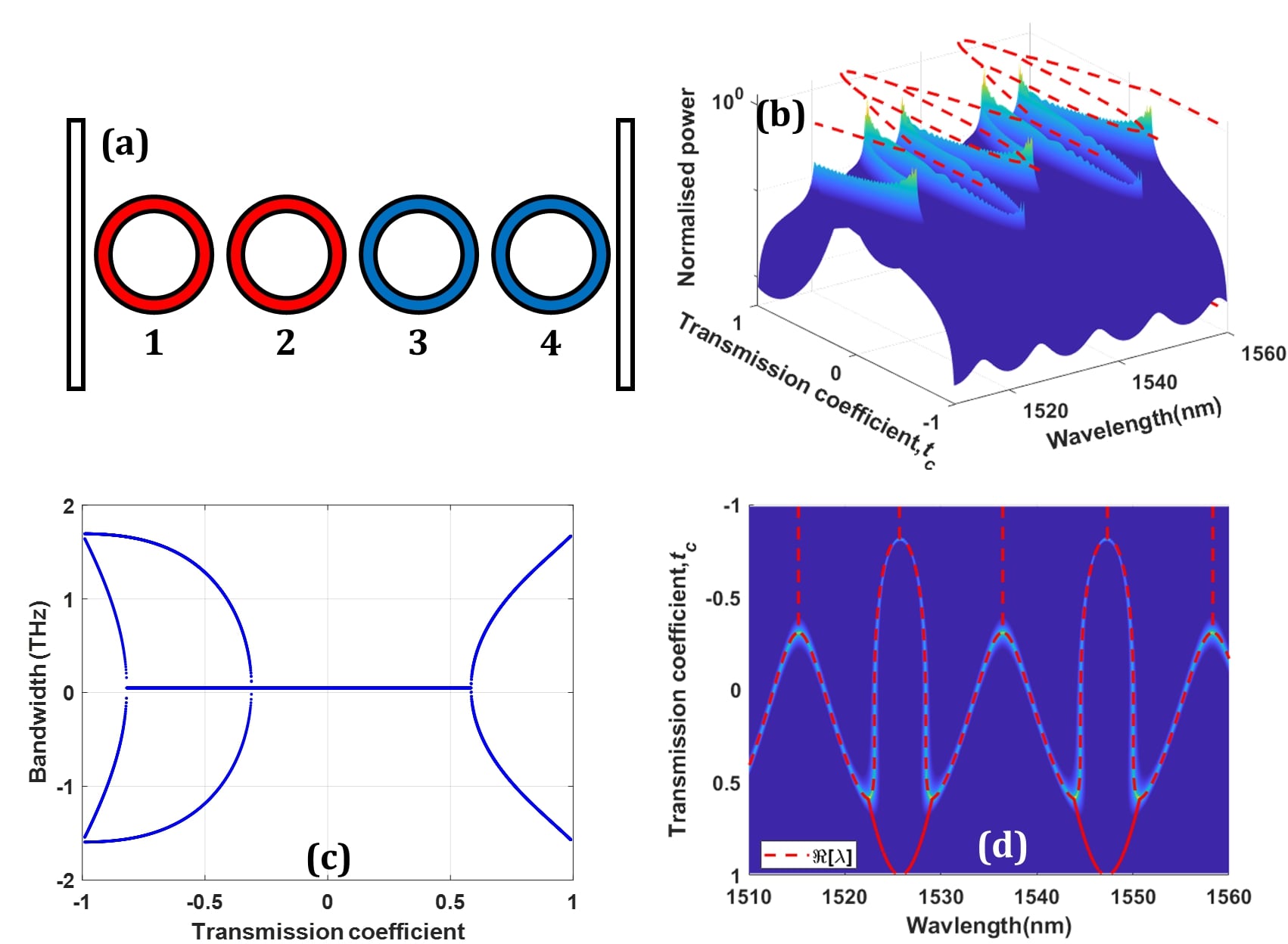}
\caption{ (a) An order $n = 4$ PT symmetric CROWe in which gain in the left-most MRR pair (red) is balanced with loss in the right-most MRR pair (blue). (b) TMM-derived 3D-map of the PT CROWe spectral response along with the real part of the eigenvalues (red dashed lines) derived from our g-TCMT model. (c) The imaginary part of the eigenvalues derived from our g-TCMT model reveals the evolution of the resonance bandwidth with the transmission coefficient, $t_c$. (d) 2D-equivalent plot of (b) illustrating the excellent agreement between the TMM-derived spectral response and evolution of the resonance positions derived from the real part of the g-TCMT eigenvalues.} \label{fig:9}
\end{figure}

As we have shown, our g-TCMT approach can provide precise derivation of the position (and indeed bandwidth) of resonance peaks in some such CROW/CROWe systems (of order $n \leq 4$) in both the weak and strong coupling regimes. We now examine a special case of an order $n = 4$ symmetric CROWe, in which gain in one pair of MRRs is balanced by loss in the other pair. Such systems can be described by non-Hermitian Hamiltonians that exhibit parity-time (PT) symmetry. That is, they are invariant under the simultaneous action of parity (space) inversion and time-reversal symmetry operators and can yield real spectra. In such systems, it is possible to identify the precise values of $t_c$ that give rise to the emergence of what are known as exceptional points (EPs), where the eigenvalues and eigenmodes coalesce, and the symmetry transitions between the unbroken and broken states. In the configuration of Fig. \ref{fig:9} (a), $R_1$ and $R_2$ (highlighted in red) are the active (gain) rings, and $R_3$ and $R_4$ (highlighted in blue) are the passive (lossy) rings, with balanced gain and loss (i.e., $\gamma_{1,2}=-\gamma_{3,4}$).

The g-TCMT derived spectra for this system are in excellent agreement with that obtained using TMM over the entire range of $t_c$. In this particular example, the spectra reveal four EPs within a single FSR; in the weak coupling regime, $0 \leq t_c\leq1$, two of these occur at the same value of $t_c$, whereas in the strong (over-coupled) regime, $-1 \leq t_c \leq 0$, the EPs occur at different values of $t_c$. Since applications of PT symmetric systems tend to focus on behaviour around the EP's, increasing their spectral density (no. of EP's in a given wavelength range) using strongly coupled CROWe's offers additional device functionality for a broad range of applications in on-chip, non-Hermitian photonics. Another peculiar feature of this 4 ring CROWe device is that it exhibits eigenvalues whose real parts are largely independent of $t_c$ over much of the unbroken PT-symmetric phase. This is different from the commonly studied two-ring PT-symmetric system, whose eigenfrequencies tend to exhibit a square-root dependence with $t_c$ in the unbroken PT state. This could be useful in applications requiring a stable frequency reference \cite{liu_far_2020}, or for laser mode-locking applications \cite{cerullo_ultrafast_2003}. Thus, here, we have demonstrated extension of the applicability of our g-TCMT model to include higher order CROWe's in which the coupling of pairs of MRRs, with balanced gain and loss, exhibit the special property of being PT symmetric.

\begin{figure}[h]
    \centering\includegraphics[width=\columnwidth]{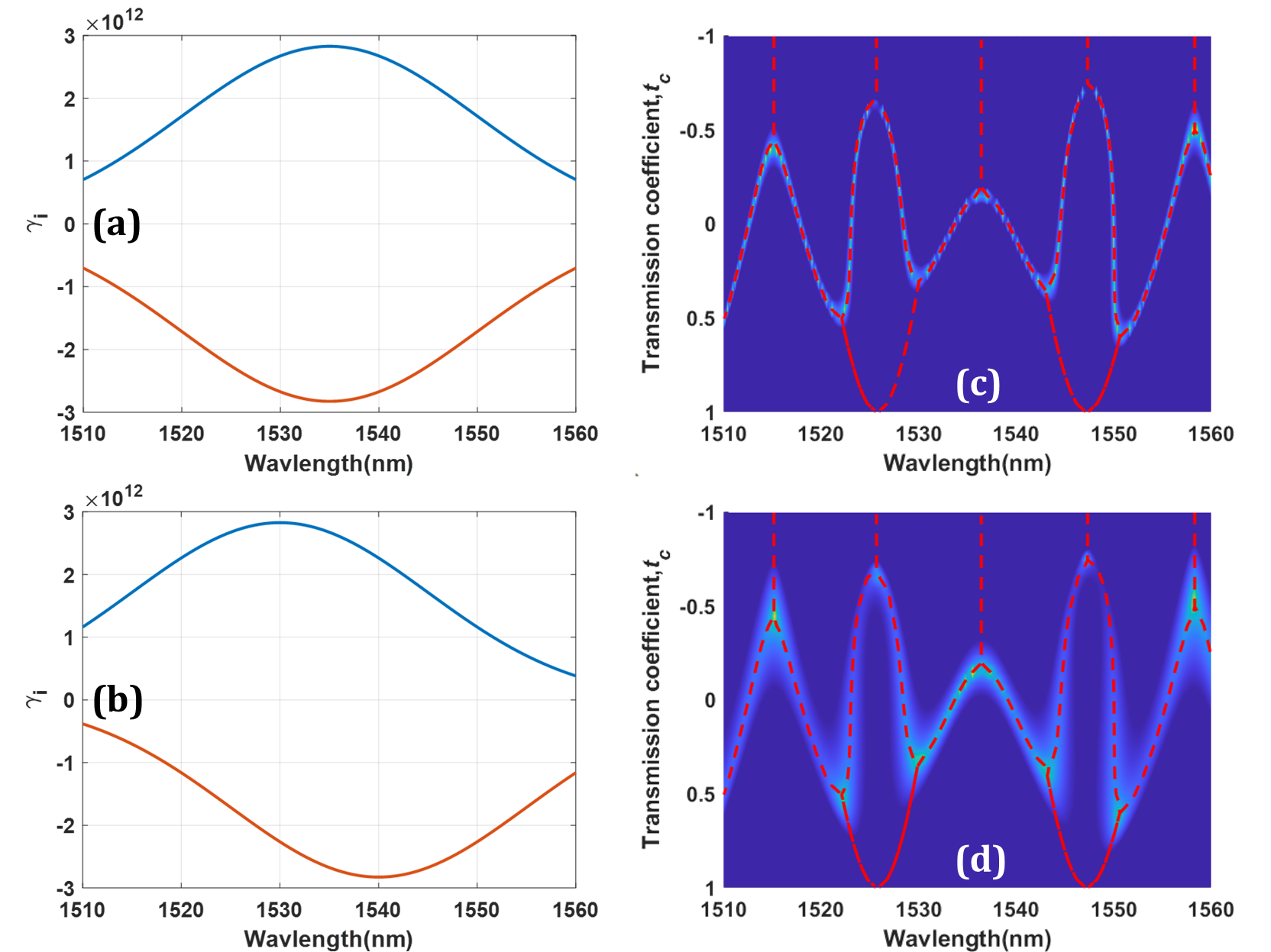}
    \caption{Spectral dependence of the gain-loss for (a) $\gamma_{1,2}=-\gamma_{3,4}$  and (b) $\gamma_{1,2} \neq -\gamma_{3,4}$ applied to the order $n = 4$ PT CROWe of Fig. \ref{fig:9}. The corresponding spectra (c, d) compares the resonance features derived by TMM with those derived from the real part of the eigenvalues (red dashed lines) using our g-TCMT model. }
    \label{fig:10}
\end{figure}

To further test the robustness of our g-TCMT model approach, specifically for the PT CROWe, we examine the effect of introducing more realistic, spectrally dependent gain/loss. Fig. \ref{fig:10} provides two such examples in which the gain and loss exhibit a spectral dependence, but in the first, Fig. \ref{fig:10} (a, c) they are equal (balanced) and in Fig. \ref{fig:10} (b, d) they are unequal.

Comparing the spectra of Fig. \ref{fig:9} (d), for which $\gamma_{1,2} = -\gamma_{3,4}$ is constant, with that of Fig. \ref{fig:10} (c), for which $\gamma_{1,2}(\lambda) = -\gamma_{3,4}(\lambda)$ we find only a small variation in the position of the EP's with $t_c$. Clearly, when $\gamma_{1,2} \neq -\gamma_{3,4}$, Fig. \ref{fig:9} (d), the most obvious feature is the increase in resonance bandwidth, although the spectral position of the EP's is largely unaffected. Critically, even when we consider more realistic spectrally-dependent gain and loss, our g-TCMT model faithfully reproduces the spectral evolution of the resonances, and the key EP features, as derived by TMM. 

\section{Conclusion}
In this paper, we have applied our recently developed g-TCMT model to serially coupled high(er) order MRR arrays, or CROW's. Our approach is to iteratively reduce coupled MRR component pairs (within the higher order CROW structure) to an equivalent single MRR pair. By studying several such higher order CROW systems, we have been able to identify the conditions necessary to ensure the method's validity, and in so doing introduced the concept of the CROWe (Coupled-Resonator Optical Waveguide with Exchange symmetry) to describe structures that satisfy these conditions.

While this paper focuses on classic MRR-based CROW's, the g-TCMT framework is not limited to this specific resonator geometry. Other types of optical resonator may also be mapped to equivalent resonator configurations, such as plane-parallel resonators, allowing for broader applicability of the method. Although g-TCMT may appear complex when applied to such higher order multi-MRR systems, it offers a powerful tool for analyzing, specifically, strongly coupled networks of resonators. In particular, because g-TCMT directly yields the system eigenvalues, it provides a natural framework for the field of non-Hermitian photonics - especially Parity-Time (PT) symmetry and exceptional point physics - that arise from such coupled resonator systems. The approach we have described represents a key advantage, e.g., over Bloch-based methods, and the standard TCMT, which are necessarily restricted to the weak coupling regime, and indeed over the TMM approach, which, despite its utility in benchmarking, requires full spectral analysis to extract similar information.

\begin{acknowledgments}
One of us (T. Li) wishes to thank the University of Manchester for the award of a doctoral scholarship. This work was supported by the Henry Royce Institute for Advanced Materials, funded through EPSRC grants EP/R00661X/1.
\end{acknowledgments}

\appendix

\section{Group theory representation of exchange symmetry}

Similar to identical particles (such as bosons), the exchange symmetry of the system can be described by a permutation group ($S_n$). According to the properties of such groups, a system composed of $n$ identical MRR's has a permutation group order of $n!$ describing its symmetry \cite{khattar_groups_2023}. Here, we take the $n = 3$ symmetric CROW system of Fig. \ref{fig:1}(a) as an example. In this case the permutation group consists of $3! = 6$ symmetric transformation elements. They are represented by cycle as: $E$(Unit element), (1 2), (2 3), (3 1), (1 2 3), (3 2 1). ($S_3$ is isomorphic to the dihedral group $D3$.) Other CROW structures with order $n = 3$, as shown in Fig. \ref{fig:3}(a, b), are merely subgroups of the $S_3$ group. As previously stated, the exchange symmetry we consider applies only to adjacent rings. Consequently, from a group-theory perspective, we are not concerned with the properties of the group itself, but solely with whether the group possesses $(j \quad j+1)$ elements. Therefore, the precise definition of exchange symmetry is that the permutation group of the system possesses a corresponding element $(j \quad j+1)$. 

Here we provide a mathematical explanation by way of proof by contradiction.
Consider an $n$-order CROW system with the component MRR's numbered, from input to output sequentially, as: 1,$\cdots$ $i$,$i+1$,$\cdots$ n, and let us assume that this system's permutation group lacks an element $(i \quad i+1)$. If, when transforming this CROW system into an equivalent resonator pair, exchange symmetry is not required, then the system may first transpose $i$ and $i+1$ into an equivalent resonator. With all other steps unchanged, since the eigenvalues of $H(\Omega_i,\Omega_{(i+1)})$ are identical to those of $H(\Omega_{(i+1)},\Omega_i)$, the final outcome will necessarily coincide with that of another CROW system of sequence 1,$\cdots$ $i+1$,$i$,$\cdots$ n. However, since in reality, these two systems exhibit different spectral responses, as illustrated in Fig. \ref{fig:3}, then this represents a contradiction. Consequently, the theory under the `equivalent resonator' assumption is confined to systems satisfying exchange symmetry.

\bibliographystyle{ieeetr}
\bibliography{apstemplate.bib}

@article{wang_generating_2021,
	title = {Generating arbitrary topological windings of a non-{Hermitian} band},
	volume = {371},
	url = {https://www.science.org/doi/10.1126/science.abf6568},
	doi = {10.1126/science.abf6568},
	number = {6535},
	urldate = {2026-05-12},
	journal = {Science},
	publisher = {American Association for the Advancement of Science},
	author = {Wang, Kai and Dutt, Avik and Yang, Ki Youl and Wojcik, Casey C. and Vučković, Jelena and Fan, Shanhui},
	month = mar,
	year = {2021},
	pages = {1240--1245},
}

@article{tsui_graphene_2020,
	title = {Graphene oxide integrated silicon photonics for detection of vapour phase volatile organic compounds},
	volume = {10},
	copyright = {© The Author(s) 2020. This work is published under http://creativecommons.org/licenses/by/4.0/ (the “License”). Notwithstanding the ProQuest Terms and Conditions, you may use this content in accordance with the terms of the License.},
	url = {https://www.proquest.com/docview/2412413495/abstract/771F2F274B2845EEPQ/1},
	doi = {10.1038/s41598-020-66389-9},
	language = {English},
	number = {1},
	urldate = {2024-06-21},
	journal = {Scientific Reports (Nature Publisher Group)},
	publisher = {Nature Publishing Group},
	author = {Tsui, H and Alsalman, Osamah and Boyang, Mao and Albrithen, Hamad and Knights, Andrew P. and Halsall, Matthew P. and Crowe, Iain F.},
	year = {2020},
	note = {Place: London, United States},
}

@incollection{khattar_groups_2023,
	address = {Cham},
	title = {Groups},
	isbn = {978-3-031-21307-6},
	url = {https://doi.org/10.1007/978-3-031-21307-6_1},
	doi = {10.1007/978-3-031-21307-6_1},
	language = {en},
	urldate = {2026-02-04},
	booktitle = {Group {Theory}},
	publisher = {Springer International Publishing},
	author = {Khattar, Dinesh and Agrawal, Neha},
	editor = {Khattar, Dinesh and Agrawal, Neha},
	year = {2023},
	pages = {1--58},
}

@article{dong_novel_2014,
	title = {Novel integration technique for silicon/{III}-{V} hybrid laser},
	volume = {22},
	copyright = {© 2014 Optical Society of America},
	issn = {1094-4087},
	url = {https://opg.optica.org/oe/abstract.cfm?uri=oe-22-22-26854},
	doi = {10.1364/OE.22.026854},
	language = {EN},
	number = {22},
	urldate = {2025-10-28},
	journal = {Optics Express},
	publisher = {Optica Publishing Group},
	author = {Dong, Po and Hu, Ting-Chen and Liow, Tsung-Yang and Chen, Young-Kai and Xie, Chongjin and Luo, Xianshu and Lo, Guo-Qiang and Kopf, Rose and Tate, Alaric},
	month = nov,
	year = {2014},
	pages = {26854--26861},
}

@article{cerullo_ultrafast_2003,
	title = {Ultrafast optical parametric amplifiers},
	volume = {74},
	issn = {0034-6748},
	url = {https://doi.org/10.1063/1.1523642},
	doi = {10.1063/1.1523642},
	number = {1},
	urldate = {2025-10-26},
	journal = {Review of Scientific Instruments},
	author = {Cerullo, Giulio and De Silvestri, Sandro},
	month = jan,
	year = {2003},
	pages = {1--18},
}

@article{liu_far_2020,
	title = {Far {Off}-{Resonance} {Laser} {Frequency} {Stabilization} {Technology}.},
	volume = {10},
	issn = {20763417},
	url = {https://go.gale.com/ps/i.do?p=AONE&sw=w&issn=20763417&v=2.1&it=r&id=GALE%7CA630945255&sid=googleScholar&linkaccess=abs},
	doi = {10.3390/app10093255},
	language = {English},
	number = {9},
	urldate = {2025-10-26},
	journal = {Applied Sciences},
	publisher = {MDPI AG},
	author = {Liu, Chang and Yue, Ziqian and Xu, Zitong and Ding, Ming and Zhai, Yueyang},
	month = may,
	year = {2020},
	pages = {NA--NA},
}

@article{leuthold_nonlinear_2010,
	title = {Nonlinear silicon photonics},
	volume = {4},
	copyright = {2010 Springer Nature Limited},
	issn = {1749-4893},
	url = {https://www.nature.com/articles/nphoton.2010.185},
	doi = {10.1038/nphoton.2010.185},
	language = {en},
	number = {8},
	urldate = {2025-10-24},
	journal = {Nature Photonics},
	publisher = {Nature Publishing Group},
	author = {Leuthold, J. and Koos, C. and Freude, W.},
	month = aug,
	year = {2010},
	pages = {535--544},
}

@article{wang_integrated_2020,
	title = {Integrated photonic quantum technologies},
	volume = {14},
	copyright = {2019 Springer Nature Limited},
	issn = {1749-4893},
	url = {https://www.nature.com/articles/s41566-019-0532-1},
	doi = {10.1038/s41566-019-0532-1},
	language = {en},
	number = {5},
	urldate = {2025-10-24},
	journal = {Nature Photonics},
	publisher = {Nature Publishing Group},
	author = {Wang, Jianwei and Sciarrino, Fabio and Laing, Anthony and Thompson, Mark G.},
	month = may,
	year = {2020},
	pages = {273--284},
}

@article{crowe_determination_2014,
	title = {Determination of the quasi-{TE} mode (in-plane) graphene linear absorption coefficient via integration with silicon-on-insulator racetrack cavity resonators},
	volume = {22},
	copyright = {© 2014 Optical Society of America},
	issn = {1094-4087},
	url = {https://opg.optica.org/oe/abstract.cfm?uri=oe-22-15-18625},
	doi = {10.1364/OE.22.018625},
	language = {EN},
	number = {15},
	urldate = {2025-10-24},
	journal = {Optics Express},
	publisher = {Optica Publishing Group},
	author = {Crowe, Iain F. and Clark, Nicholas and Hussein, Siham and Towlson, Brian and Whittaker, Eric and Milosevic, Milan M. and Gardes, Frederic Y. and Mashanovich, Goran Z. and Halsall, Matthew P. and Vijayaraghaven, Aravind},
	month = jul,
	year = {2014},
	pages = {18625--18632},
}

@article{hussein_raman_2017,
	title = {Raman {Mapping} {Analysis} of {Graphene}-{Integrated} {Silicon} {Micro}-{Ring} {Resonators}},
	volume = {12},
	issn = {1556-276X},
	url = {https://doi.org/10.1186/s11671-017-2374-4},
	doi = {10.1186/s11671-017-2374-4},
	language = {en},
	number = {1},
	urldate = {2025-10-24},
	journal = {Nanoscale Research Letters},
	author = {Hussein, Siham M. and Crowe, Iain F. and Clark, Nick and Milosevic, Milan and Vijayaraghavan, Aravind and Gardes, Frederic Y. and Mashanovich, Goran Z. and Halsall, Matthew P.},
	month = nov,
	year = {2017},
	pages = {600},
}

@article{littlejohns_cornerstones_2020,
	title = {{CORNERSTONE}’s {Silicon} {Photonics} {Rapid} {Prototyping} {Platforms}: {Current} {Status} and {Future} {Outlook}},
	volume = {10},
	copyright = {http://creativecommons.org/licenses/by/3.0/},
	issn = {2076-3417},
	shorttitle = {{CORNERSTONE}’s {Silicon} {Photonics} {Rapid} {Prototyping} {Platforms}},
	url = {https://www.mdpi.com/2076-3417/10/22/8201},
	doi = {10.3390/app10228201},
	language = {en},
	number = {22},
	urldate = {2025-10-24},
	journal = {Applied Sciences},
	publisher = {Multidisciplinary Digital Publishing Institute},
	author = {Littlejohns, Callum G. and Rowe, David J. and Du, Han and Li, Ke and Zhang, Weiwei and Cao, Wei and Dominguez Bucio, Thalia and Yan, Xingzhao and Banakar, Mehdi and Tran, Dehn and Liu, Shenghao and Meng, Fanfan and Chen, Bigeng and Qi, Yanli and Chen, Xia and Nedeljkovic, Milos and Mastronardi, Lorenzo and Maharjan, Rijan and Bohora, Sanket and Dhakal, Ashim and Crowe, Iain and Khurana, Ankur and Balram, Krishna C. and Zagaglia, Luca and Floris, Francesco and O’Brien, Peter and Di Gaetano, Eugenio and Chong, Harold M. H. and Gardes, Frederic Y. and Thomson, David J. and Mashanovich, Goran Z. and Sorel, Marc and Reed, Graham T.},
	month = jan,
	year = {2020},
	pages = {8201},
}

@article{xu_propagation_2000,
	title = {Propagation and second-harmonic generation of electromagnetic waves in a coupled-resonator optical waveguide},
	volume = {17},
	copyright = {© 2000 Optical Society of America},
	issn = {1520-8540},
	url = {https://opg.optica.org/josab/abstract.cfm?uri=josab-17-3-387},
	doi = {10.1364/JOSAB.17.000387},
	language = {EN},
	number = {3},
	urldate = {2025-10-22},
	journal = {JOSA B},
	publisher = {Optica Publishing Group},
	author = {Xu, Youg and Lee, Reginald K. and Yariv, Amnon},
	month = mar,
	year = {2000},
	pages = {387--400},
}

@article{ji_microring-resonator-based_2011,
	title = {Microring-resonator-based four-port optical router for photonic networks-on-chip},
	volume = {19},
	copyright = {© 2011 OSA},
	issn = {1094-4087},
	url = {https://opg.optica.org/oe/abstract.cfm?uri=oe-19-20-18945},
	doi = {10.1364/OE.19.018945},
	language = {EN},
	number = {20},
	urldate = {2025-10-22},
	journal = {Optics Express},
	publisher = {Optica Publishing Group},
	author = {Ji, Ruiqiang and Yang, Lin and Zhang, Lei and Tian, Yonghui and Ding, Jianfeng and Chen, Hongtao and Lu, Yangyang and Zhou, Ping and Zhu, Weiwei},
	month = sep,
	year = {2011},
	pages = {18945--18955},
}

@article{yuan_5_2024,
	title = {A 5 × 200 {Gbps} microring modulator silicon chip empowered by two-segment {Z}-shape junctions},
	volume = {15},
	copyright = {2024 The Author(s)},
	issn = {2041-1723},
	url = {https://www.nature.com/articles/s41467-024-45301-3},
	doi = {10.1038/s41467-024-45301-3},
	language = {en},
	number = {1},
	urldate = {2025-10-21},
	journal = {Nature Communications},
	publisher = {Nature Publishing Group},
	author = {Yuan, Yuan and Peng, Yiwei and Sorin, Wayne V. and Cheung, Stanley and Huang, Zhihong and Liang, Di and Fiorentino, Marco and Beausoleil, Raymond G.},
	month = jan,
	year = {2024},
	pages = {918},
}

@article{li_n-order_2024,
	title = {\textit{{N}}-order generalized-temporal coupled mode theory (g-{TCMT}) model: extending the spectral range for arbitrary coupling of optical resonators and application in {Parity}-{Time} ({PT}) symmetry},
	volume = {32},
	issn = {1094-4087},
	shorttitle = {\textit{{N}}-order generalized-temporal coupled mode theory (g-{TCMT}) model},
	url = {https://opg.optica.org/oe/abstract.cfm?uri=oe-32-26-46569},
	doi = {10.1364/OE.544255},
	language = {EN},
	number = {26},
	urldate = {2025-09-24},
	journal = {Optics Express},
	publisher = {Optica Publishing Group},
	author = {Li, Tianrui and Halsall, Matthew P. and Crowe, Iain F.},
	month = dec,
	year = {2024},
	pages = {46569--46577},
}

@article{ruter_observation_2010,
	title = {Observation of parity–time symmetry in optics},
	volume = {6},
	copyright = {2010 Springer Nature Limited},
	issn = {1745-2481},
	url = {https://www.nature.com/articles/nphys1515},
	doi = {10.1038/nphys1515},
	language = {en},
	number = {3},
	urldate = {2023-11-22},
	journal = {Nature Physics},
	publisher = {Nature Publishing Group},
	author = {Rüter, Christian E. and Makris, Konstantinos G. and El-Ganainy, Ramy and Christodoulides, Demetrios N. and Segev, Mordechai and Kip, Detlef},
	month = mar,
	year = {2010},
	note = {Number: 3},
	pages = {192--195},
}

@article{ong_ultra-high-contrast_2013,
	title = {Ultra-{High}-{Contrast} and {Tunable}-{Bandwidth} {Filter} {Using} {Cascaded} {High}-{Order} {Silicon} {Microring} {Filters}},
	volume = {25},
	issn = {1941-0174},
	url = {https://ieeexplore.ieee.org/document/6527974},
	doi = {10.1109/LPT.2013.2267539},
	number = {16},
	urldate = {2025-04-03},
	journal = {IEEE Photonics Technology Letters},
	author = {Ong, Jun Rong and Kumar, Ranjeet and Mookherjea, Shayan},
	month = aug,
	year = {2013},
	note = {Conference Name: IEEE Photonics Technology Letters},
	pages = {1543--1546},
}

@article{de_aguiar_automatic_2019,
	title = {Automatic {Tuning} of {Silicon} {Photonics} {Microring} {Filter} {Array} for {Hitless} {Reconfigurable} {Add}–{Drop}},
	volume = {37},
	issn = {1558-2213},
	url = {https://ieeexplore.ieee.org/document/8713385},
	doi = {10.1109/JLT.2019.2916473},
	number = {16},
	urldate = {2025-04-03},
	journal = {Journal of Lightwave Technology},
	author = {de Aguiar, Douglas Oliveira Morais and Milanizadeh, Maziyar and Guglielmi, Emanuele and Zanetto, Francesco and Ferrari, Giorgio and Sampietro, Marco and Morichetti, Francesco and Melloni, Andrea},
	month = aug,
	year = {2019},
	note = {Conference Name: Journal of Lightwave Technology},
	pages = {3939--3947},
}

@article{yi_highly_2010,
	title = {Highly sensitive silicon microring sensor with sharp asymmetrical resonance},
	volume = {18},
	copyright = {© 2010 OSA},
	issn = {1094-4087},
	url = {https://opg.optica.org/oe/abstract.cfm?uri=oe-18-3-2967},
	doi = {10.1364/OE.18.002967},
	language = {EN},
	number = {3},
	urldate = {2025-04-03},
	journal = {Optics Express},
	publisher = {Optica Publishing Group},
	author = {Yi, Huaxiang and Citrin, D. S. and Zhou, Zhiping},
	month = feb,
	year = {2010},
	pages = {2967--2972},
}

@article{youplao_microring_2018,
	title = {Microring stereo sensor model using {Kerr}–{Vernier} effect for bio-cell sensor and communication},
	volume = {17},
	issn = {1878-7789},
	url = {https://www.sciencedirect.com/science/article/pii/S1878778918300759},
	doi = {10.1016/j.nancom.2018.06.002},
	urldate = {2025-04-03},
	journal = {Nano Communication Networks},
	author = {Youplao, P. and Pornsuwancharoen, N. and Amiri, I. S. and Jalil, M. A. and Aziz, M. S. and Ali, J. and Singh, G. and Yupapin, P. and Grattan, K. T. V.},
	month = sep,
	year = {2018},
	pages = {30--35},
}

@misc{wang_fully_2021,
	title = {Fully symmetric controllable integrated three-resonator photonic molecule},
	url = {http://arxiv.org/abs/2105.10815},
	doi = {10.48550/arXiv.2105.10815},
	urldate = {2025-04-03},
	publisher = {arXiv},
	author = {Wang, Jiawei and Liu, Kaikai and Zhao, Qiancheng and Isichenko, Andrei and Rudy, Ryan Q. and Blumenthal, Daniel J.},
	month = may,
	year = {2021},
	note = {arXiv:2105.10815 [physics]},
}

@misc{noauthor_polymer_nodate,
	title = {Polymer microring coupled-resonator optical waveguides {\textbar} {IEEE} {Journals} \& {Magazine} {\textbar} {IEEE} {Xplore}},
	url = {https://ieeexplore-ieee-org.manchester.idm.oclc.org/document/1618774/metrics#metrics},
	urldate = {2025-04-03},
}

@article{yariv_coupled-resonator_1999,
	title = {Coupled-resonator optical waveguide:?a proposal and analysis},
	volume = {24},
	copyright = {© 1999 Optical Society of America},
	issn = {1539-4794},
	shorttitle = {Coupled-resonator optical waveguide},
	url = {https://opg.optica.org/ol/abstract.cfm?uri=ol-24-11-711},
	doi = {10.1364/OL.24.000711},
	language = {EN},
	number = {11},
	urldate = {2025-04-03},
	journal = {Optics Letters},
	publisher = {Optica Publishing Group},
	author = {Yariv, Amnon and Xu, Yong and Lee, Reginald K. and Scherer, Axel},
	month = jun,
	year = {1999},
	pages = {711--713},
}

@article{poon_matrix_2004,
	title = {Matrix analysis of microring coupled-resonator optical waveguides},
	volume = {12},
	copyright = {© 2004 Optical Society of America},
	issn = {1094-4087},
	url = {https://opg.optica.org/oe/abstract.cfm?uri=oe-12-1-90},
	doi = {10.1364/OPEX.12.000090},
	language = {EN},
	number = {1},
	urldate = {2025-04-03},
	journal = {Optics Express},
	publisher = {Optica Publishing Group},
	author = {Poon, Joyce K. S. and Scheuer, Jacob and Mookherjea, Shayan and Paloczi, George T. and Huang, Yanyi and Yariv, Amnon},
	month = jan,
	year = {2004},
	pages = {90--103},
}

@article{poon_transmission_2006,
	title = {Transmission and group delay of microring coupled-resonator optical waveguides},
	volume = {31},
	copyright = {\&\#169; 2006 Optical Society of America},
	issn = {1539-4794},
	url = {https://opg.optica.org/ol/abstract.cfm?uri=ol-31-4-456},
	doi = {10.1364/OL.31.000456},
	language = {EN},
	number = {4},
	urldate = {2025-04-02},
	journal = {Optics Letters},
	publisher = {Optica Publishing Group},
	author = {Poon, Joyce K. and Zhu, Lin and DeRose, Guy A. and Yariv, Amnon},
	month = feb,
	year = {2006},
	pages = {456--458},
}

@article{peng_paritytime-symmetric_2014,
	title = {Parity–time-symmetric whispering-gallery microcavities},
	volume = {10},
	issn = {1745-2481},
	url = {https://doi.org/10.1038/nphys2927},
	doi = {10.1038/nphys2927},
	number = {5},
	journal = {Nature Physics},
	author = {Peng, Bo and Özdemir, Şahin Kaya and Lei, Fuchuan and Monifi, Faraz and Gianfreda, Mariagiovanna and Long, Gui Lu and Fan, Shanhui and Nori, Franco and Bender, Carl M. and Yang, Lan},
	month = may,
	year = {2014},
	pages = {394--398},
}

@article{chen_exceptional_2017,
	title = {Exceptional points enhance sensing in an optical microcavity},
	volume = {548},
	issn = {1476-4687},
	url = {https://doi.org/10.1038/nature23281},
	doi = {10.1038/nature23281},
	number = {7666},
	journal = {Nature},
	author = {Chen, Weijian and Kaya Özdemir, Şahin and Zhao, Guangming and Wiersig, Jan and Yang, Lan},
	month = aug,
	year = {2017},
	pages = {192--196},
}

@article{dai_highly_2009,
	title = {Highly sensitive sensor based on an ultra-high-{Q} {Mach}-{Zehnder} interferometer-coupled microring},
	volume = {26},
	copyright = {© 2009 Optical Society of America},
	issn = {1520-8540},
	url = {https://opg.optica.org/josab/abstract.cfm?uri=josab-26-3-511},
	doi = {10.1364/JOSAB.26.000511},
	language = {EN},
	number = {3},
	urldate = {2024-09-24},
	journal = {JOSA B},
	publisher = {Optica Publishing Group},
	author = {Dai, Daoxin and He, Sailing},
	month = mar,
	year = {2009},
	pages = {511--516},
}

@article{popovic_coupling-induced_2006,
	title = {Coupling-induced resonance frequency shifts in coupled dielectric multi-cavity filters},
	volume = {14},
	copyright = {© 2006 Optical Society of America},
	issn = {1094-4087},
	url = {https://opg.optica.org/oe/abstract.cfm?uri=oe-14-3-1208},
	doi = {10.1364/OE.14.001208},
	language = {EN},
	number = {3},
	urldate = {2024-08-12},
	journal = {Optics Express},
	publisher = {Optica Publishing Group},
	author = {Popović, Miloš A. and Manolatou, Christina and Watts, Michael R.},
	month = feb,
	year = {2006},
	pages = {1208--1222},
}

@article{tan_high-order_2013,
	title = {High-order all-optical differential equation solver based on microring resonators},
	volume = {38},
	copyright = {© 2013 Optical Society of America},
	issn = {1539-4794},
	url = {https://opg.optica.org/ol/abstract.cfm?uri=ol-38-19-3735},
	doi = {10.1364/OL.38.003735},
	language = {EN},
	number = {19},
	urldate = {2024-02-20},
	journal = {Optics Letters},
	publisher = {Optica Publishing Group},
	author = {Tan, Sisi and Xiang, Lei and Zou, Jinghui and Zhang, Qiang and Wu, Zhao and Yu, Yu and Dong, Jianji and Zhang, Xinliang},
	month = oct,
	year = {2013},
	pages = {3735--3738},
}

@article{sun_induced_2021,
	title = {Induced homomorphism: {Kirchhoff}’s law in photonics},
	volume = {10},
	copyright = {© 2021. This work is published under http://creativecommons.org/licenses/by/4.0 (the “License”). Notwithstanding the ProQuest Terms and Conditions, you may use this content in accordance with the terms of the License.},
	issn = {21928606},
	shorttitle = {Induced homomorphism},
	url = {https://www.proquest.com/docview/2515813598/abstract/56282E4C044B4E61PQ/1},
	doi = {10.1515/nanoph-2020-0655},
	language = {English},
	number = {6},
	urldate = {2024-02-06},
	journal = {Nanophotonics},
	publisher = {Walter de Gruyter GmbH},
	author = {Sun, Shuai and Miscuglio, Mario and Ma, Xiaoxuan and Ma, Zhizhen and Chen, Shen and Kayraklioglu, Engin and Anderson, Jeffery and Ghazawi, Tarek El and Sorger, Volker J.},
	year = {2021},
	note = {Num Pages: 1711-1721},
	pages = {1711--1721},
}

@article{ye_reconfigurable_2024,
	title = {Reconfigurable application-specific photonic integrated circuit for solving partial differential equations},
	copyright = {De Gruyter expressly reserves the right to use all content for commercial text and data mining within the meaning of Section 44b of the German Copyright Act.},
	issn = {2192-8614},
	url = {https://www.degruyter.com/document/doi/10.1515/nanoph-2023-0732/html},
	doi = {10.1515/nanoph-2023-0732},
	language = {en},
	urldate = {2024-02-05},
	journal = {Nanophotonics},
	publisher = {De Gruyter},
	author = {Ye, Jiachi and Shen, Chen and Peserico, Nicola and Meng, Jiawei and Ma, Xiaoxuan and Nouri, Behrouz Movahhed and Popescu, Cosmin-Constantin and Hu, Juejun and Kang, Haoyan and Wang, Hao and El-Ghazawi, Tarek and Dalir, Hamed and Sorger, Volker J.},
	month = jan,
	year = {2024},
}

@article{prabhu_extreme_2010,
	title = {Extreme {Miniaturization} of {Silicon} {Add}–{Drop} {Microring} {Filters} for {VLSI} {Photonics} {Applications}},
	volume = {2},
	issn = {1943-0655},
	url = {http://ieeexplore.ieee.org/document/5460960/},
	doi = {10.1109/JPHOT.2010.2049831},
	language = {en},
	number = {3},
	urldate = {2023-11-28},
	journal = {IEEE Photonics Journal},
	author = {Prabhu, Ashok M and Tsay, Alan and {Zhanghua Han} and {Vien Van}},
	month = jun,
	year = {2010},
	pages = {436--444},
}

@article{zhao_paritytime_2018,
	title = {Parity–time symmetric photonics},
	volume = {5},
	issn = {2095-5138},
	url = {https://doi.org/10.1093/nsr/nwy011},
	doi = {10.1093/nsr/nwy011},
	number = {2},
	urldate = {2023-11-24},
	journal = {National Science Review},
	author = {Zhao, Han and Feng, Liang},
	month = mar,
	year = {2018},
	pages = {183--199},
}

@article{rabiei_polymer_2002,
	title = {Polymer micro-ring filters and modulators},
	volume = {20},
	issn = {0733-8724},
	url = {http://ieeexplore.ieee.org/document/1161608/},
	doi = {10.1109/JLT.2002.803058},
	language = {en},
	number = {11},
	urldate = {2023-11-23},
	journal = {Journal of Lightwave Technology},
	author = {Rabiei, P. and Steier, W.H. and {Cheng Zhang} and Dalton, L.R.},
	month = nov,
	year = {2002},
	pages = {1968--1975},
}

@inproceedings{ren_pt-symmetric_2018,
	title = {{PT}-{Symmetric} {Microring} {Laser} {Gyroscope}},
	url = {https://ieeexplore.ieee.org/document/8426615},
	urldate = {2023-11-23},
	booktitle = {2018 {Conference} on {Lasers} and {Electro}-{Optics} ({CLEO})},
	author = {Ren, J. and Harari, G. and Hassan, A. U. and Chow, W. and Soltani, M. and Hokmabadi, M. P. and Christodoulides, D. and Khajavikhan, M.},
	month = may,
	year = {2018},
	pages = {1--2},
}

@article{zhang_bandwidth_2022,
	title = {Bandwidth {Tunable} {Optical} {Bandpass} {Filter} {Based} on {Parity}-{Time} {Symmetry}},
	volume = {13},
	copyright = {http://creativecommons.org/licenses/by/3.0/},
	issn = {2072-666X},
	url = {https://www.mdpi.com/2072-666X/13/1/89},
	doi = {10.3390/mi13010089},
	language = {en},
	number = {1},
	urldate = {2023-11-22},
	journal = {Micromachines},
	publisher = {Multidisciplinary Digital Publishing Institute},
	author = {Zhang, Bowen and Chen, Nuo and Lu, Xinda and Hu, Yuhang and Yang, Zihao and Zhang, Xinliang and Xu, Jing},
	month = jan,
	year = {2022},
	note = {Number: 1},
	pages = {89},
}

@article{liu_integrated_2017,
	title = {An integrated parity-time symmetric wavelength-tunable single-mode microring laser},
	volume = {8},
	issn = {2041-1723},
	url = {https://www.nature.com/articles/ncomms15389},
	doi = {10.1038/ncomms15389},
	language = {en},
	number = {1},
	urldate = {2023-11-22},
	journal = {Nature Communications},
	author = {Liu, Weilin and Li, Ming and Guzzon, Robert S. and Norberg, Erik J. and Parker, John S. and Lu, Mingzhi and Coldren, Larry A. and Yao, Jianping},
	month = may,
	year = {2017},
	pages = {15389},
}

\end{document}